\documentclass[10pt,aps,prl,twocolumn,groupedaddress,amsmath]{revtex4-2}

\usepackage{graphicx}
\usepackage{booktabs}

\begin{document}


\title{Capturing thermal effects beyond the zero-temperature approximation using the uniform electron gas}


\author{Brianna Aguilar-Solis}

\author{Brittany P. Harding}

\author{Aurora Pribram-Jones}
\email[]{apj@ucmerced.edu}
\affiliation{University of California, Merced, 5200 North Lake Road, Merced, CA 95343, USA}


\date{\today}

\begin{abstract}
Density functional theory at finite temperatures often relies on the zero-temperature approximation, which uses a ground-state exchange-correlation functional with thermalized densities. This approach, however, neglects the explicit temperature dependence of the exchange-correlation free energy -- a key factor in regimes such as warm dense matter, where both electronic and thermal effects are significant. In this work, we introduce the entropy-corrected zero-temperature approach, in which the exchange-correlation entropy is extracted using the generalized thermal adiabatic connection formula to construct a thermal correction to the standard zero-temperature approximation. Using a uniform electron gas parametrization, we compare this approach to the finite-temperature adiabatic connection and demonstrate that it performs best at lower densities. This provides a useful complement to zero-temperature density functional approximations, which generally perform better at moderate-to-large densities.  We further identify a density-dependent intersection between the adiabatic connection curves, revealing a dependence on the ground state correlation energy and correlation potential. Additionally, extension of the entropy corrected approach applied as a local density approximation--like temperature correction to the zero temperature approximation are discussed. 
\begin{center}
\includegraphics[width=0.7\columnwidth]{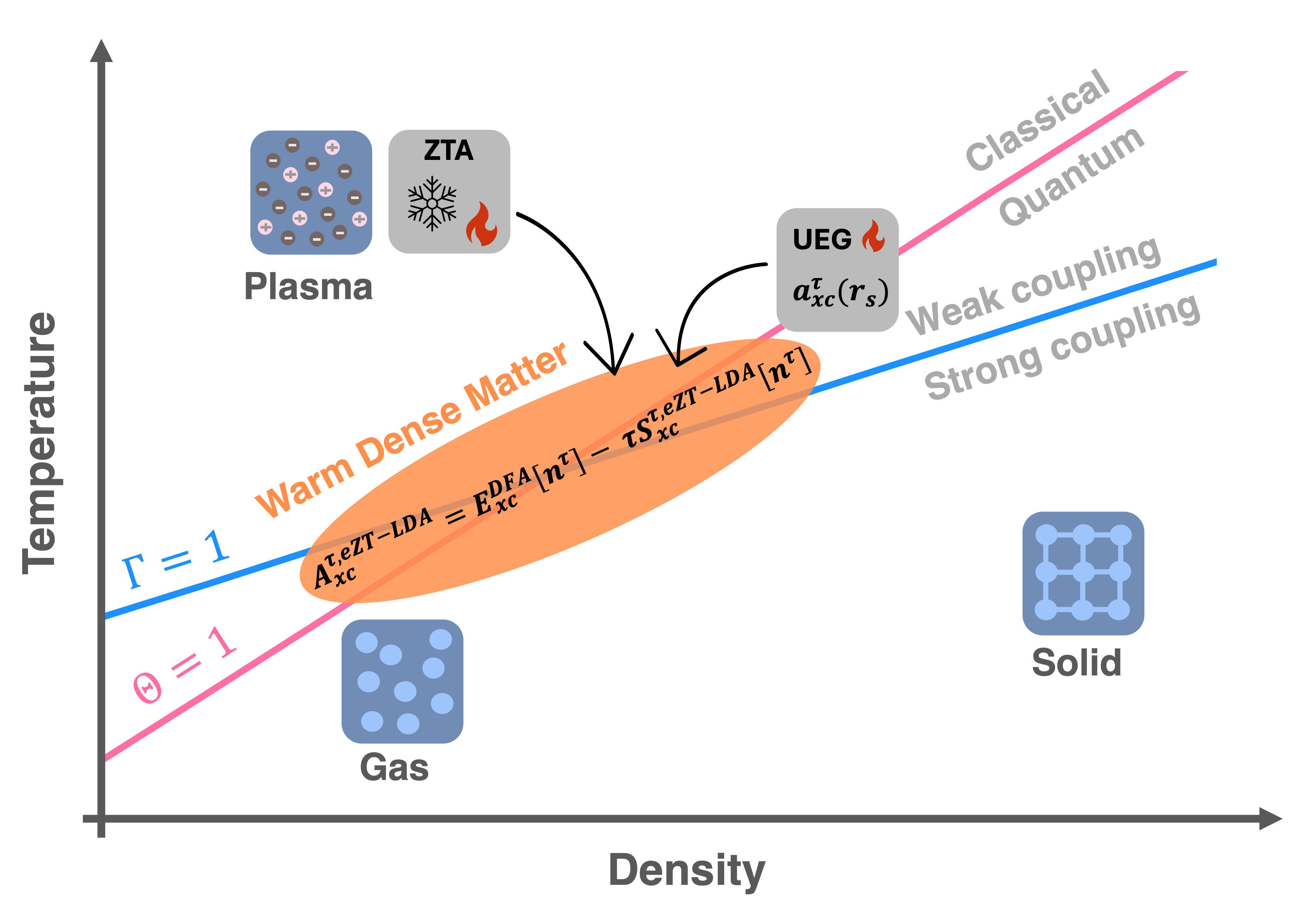}
\end{center}

\end{abstract}

\maketitle

\section{\label{introduction}Introduction} 

The many-body problem in quantum mechanics remains one of the central challenges in modern physics. Numerous methods have been developed in an attempt to simplify this problem while still capturing the complex physics of interacting  particles \cite{SO82}. Among these, density functional theory (DFT) \cite{HK64,KS65} has become one of the most widely used and successful approaches in condensed matter physics and quantum chemistry \cite{B12,PY89}. DFT provides an exact framework in which the many-body problem is mapped onto a non-interacting reference that, in principle, returns the exact energy of the fully interacting system. While DFT successfully describes a wide range of ground-state properties, this formulation neglects temperature effects when applied to temperature-dependent systems, like those in the warm dense matter (WDM) regime \cite{GDRT14}. Understanding how DFT performs, and how it can be extended, in this intermediate region is essential for accurately modeling these conditions. 

WDM is a transitional regime between condensed matter and plasma physics. It is found naturally in planetary cores \cite{BMRV14,NRB08} and can also be generated experimentally along the path to ignition in inertial confinement fusion (ICF) experiments, such as those conducted at the National Ignition Facility at Lawrence Livermore National Laboratory \cite{MHVT08,MSAA05,AAA24}. Gaining a better understanding of WDM can help guide target fabrication in ICF experiments, bringing us closer to utilizing fusion energy as a clean, renewable energy source. The WDM regime is characterized by strong electron correlation, thermal effects, and quantum mechanical behavior of electrons. Additionally, the electron degeneracy parameter, $\Theta = k_BT/E_F$, is close to unity in this regime \cite{GDRT14}, lying between the fully degenerate, $\Theta << 1$, and semi-classical, $\Theta >> 1$, limits. The Coulomb coupling parameter, $\Gamma$, is likewise close to one in this region, reflecting the simultaneous importance of quantum and correlation effects. As such, modeling methods typically used in plasma physics or condensed matter physics, like ground-state DFT, struggle to accurately describe this regime \cite{GDRT14,DGB18,KCT16}.

To bridge this gap, thermal DFT \cite{HK64,KS65,M65,SP88} extends the ground-state formalism to finite temperatures, allowing for explicit temperature dependence. Accurate approximations to the exchange-correlation (XC) free energy are necessary for studying temperature-dependent systems like WDM \cite{KCT16, GDSM17,BDHC13}. A common approach for including temperature dependence in DFT involves using a ground-state XC functional with thermally weighted densities, which accounts for the temperature dependence of the density. This is known as the zero-temperature approximation (ZTA) \cite{BSGP15}. While this approximation captures the implicit part of the temperature dependence, it neglects the explicit temperature dependence of the XC free energy and fails to reproduce accurate results at high temperatures ({\em i.e.}, those exceeding 10,000 K \cite{SD14}).  Work by Karasiev and co-workers has also shown the importance of considering the explicit temperature dependence at various densities and temperatures within the WDM regime \cite{KCT16}. 

The adiabatic connection formula (ACF) is an exact formalism that is used to analyze and build approximations to the XC energy in the ground state \cite{HG75,LP75,GL76, H84}. In finite-temperature systems, an adapted version, known as the finite-temperature adiabatic connection (FTAC), is used to analyze and approximate XC free energies \cite{PPFS11,PB16,HMP22}.
The generalized thermal adiabatic connection (GTAC)  \cite{HMXP24} is an additional thermalized version of the ACF that allows for controlled variations in both interaction strength and a fictitious temperature parameter, while preserving relative scaling relations and other formal conditions established by the FTAC. The GTAC enables extraction of the XC entropy, which we apply in this work to construct the entropy-corrected zero-temperature (eZT) approach -- a thermal correction to the ZTA. The derived eZT is implemented using a parametrization for the uniform electron gas (UEG) \cite{GDSM17} and compared to the FTAC. The resulting framework provides a Local Density Approximation (LDA) -like correction that can be used as a post-hoc correction to ZTA calculations. As such, this offers a more complete approach for modeling WDM without sacrificing the useful accuracy of beyond-LDA zero-temperature XC density functional approximations (DFA). 

\section{\label{methods}Methods} 

One of the most successful approaches to solving the many-body problem is Kohn-Sham density functional theory (KS-DFT) \cite{KS65}. This method introduces a non-interacting reference system that reproduces the exact ground-state energy of the fully interacting system. In practice, the Kohn-Sham equations,

\begin{gather}
\label{eq:ks_eqns}
\left\{-\frac{1}{2}\nabla^2 +v_\mathrm{s}(\mathbf{r})\right\}\phi_i(\mathbf{r}) = \epsilon_i\phi_i(\mathbf{r}) ~, \notag \\
n(\mathbf{r}) = \sum^N_{i=1} |\phi_i(\mathbf{r})|^2~,
\end{gather} 

\noindent where $v_\mathrm{s}$ is the KS potential, $\phi$ is the KS orbitals and $\epsilon$ is the energy, are solved iteratively to determine the total energy as a functional of the density:

\begin{equation}
\label{eq:totalE}
E[n] = T_\mathrm{s}[n] + V_\mathrm{ext}[n]+U_\mathrm{H}[n]+E_\mathrm{xc}[n]~.
\end{equation} 

$T_\mathrm{s}$ is the non-interacting kinetic energy, $U_\mathrm{H}$ is the classical Coulomb repulsion, $V_\mathrm{ext}$ is the external potential, and $E_\mathrm{xc}$ is the exchange-correlation energy. While this method, in principle, yields the exact energy of the fully interacting system, in practice, approximations are made to $E_\mathrm{xc}[n]$, as the analytical form is not known. Although this is a relatively small component of the total energy, it plays a critical role in describing bonding and capturing the quantum mechanical behavior of electrons \cite{Teller62,KP00}. As such, it is important to build accurate approximations for $E_\mathrm{xc}$. The ACF provides an exact formalism for the exchange-correlation energy that can be used to build approximations to $E_\mathrm{xc}$ and facilitate analysis of exchange and correlation effects. 

The ground-state ACF gives an exact expression for the exchange-correlation energy at zero temperature,

\begin{equation}
\label{eq:exc}
E_\mathrm{{xc}}[n] = \int_{0}^{1} d\lambda\frac{U_\mathrm{{xc}}^{\lambda}[n]}{\lambda}.
\end{equation} 

\begin{figure}[ht]
\centering
\includegraphics[width= 1\columnwidth]{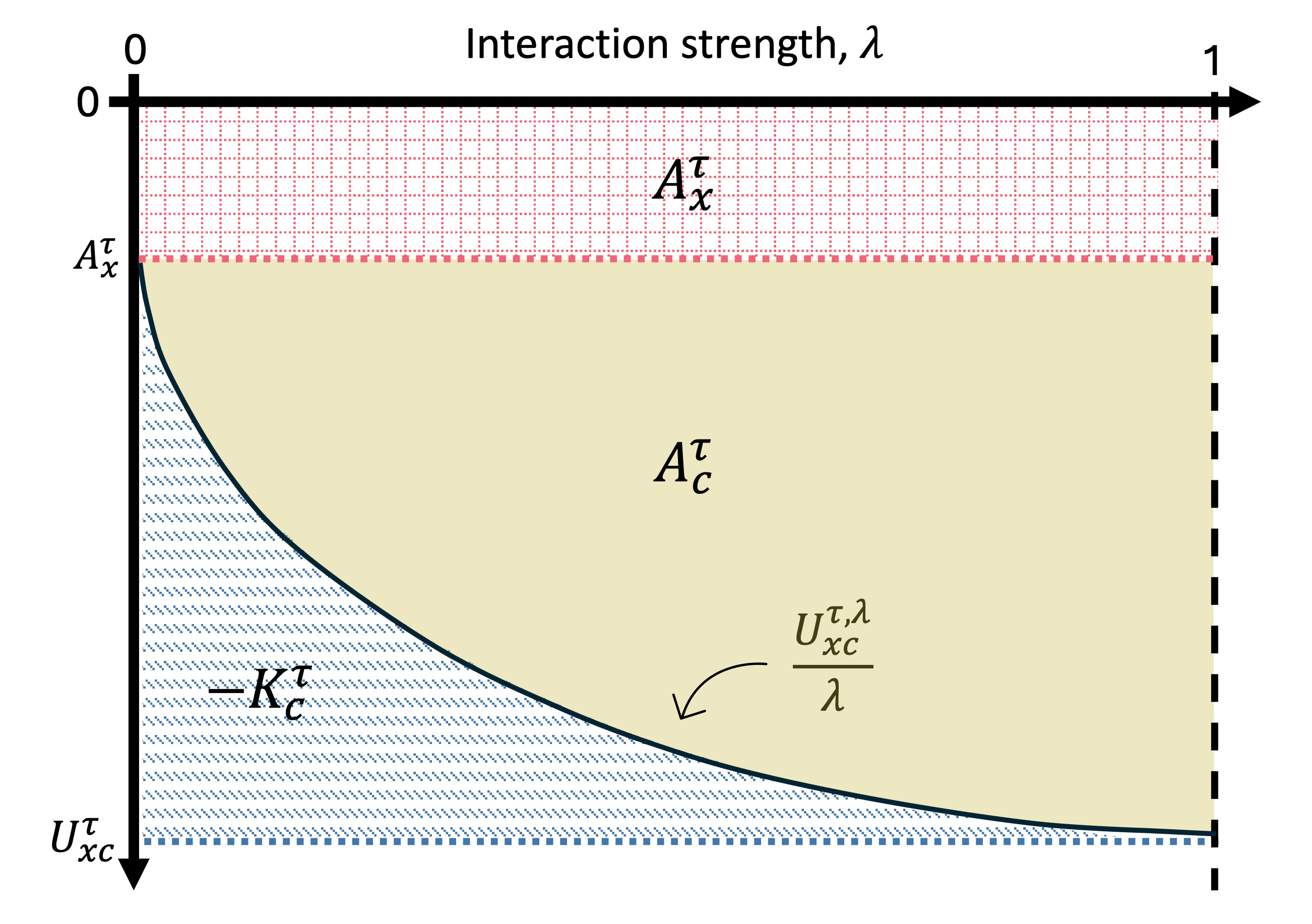}
\caption{\label{fig:ftac_updated} Geometric interpretation of the FTAC \cite{PPFS11,HMP22}.}
\end{figure}

\noindent In this expression, $\lambda$ scales the electron-electron interaction between the KS system ($\lambda = 0$) and the fully interacting, ``real" system ($\lambda = 1$). This is done while holding the density, $n(\mathbf{r})$, constant for the potential XC functional evaluated at interaction strength $\lambda$, $U_\mathrm{xc}^\lambda[n]$. The ACF integrand is formed by scaling the universal functional by $\lambda$, $F^{\lambda}[n] = \min_{\Psi \rightarrow n}\langle \Psi | \hat{T}+\lambda\hat{V}_\mathrm{ee} | \Psi \rangle$. Where $\Psi^\lambda$ is the minimizing wavefunction for a given $\lambda$. The ACF provides a geometric interpretation of exchange and correlation components, as shown in Figure \ref{fig:ftac_updated} for the finite-temperature ACF, and helps us verify that functional approximations obey known constraints on the exchange-correlation energy \cite{PRTS05}.

Moving to a temperature-dependent framework involves solving the Mermin-Kohn-Sham equations,

\begin{gather}
\label{eq:mks_eqns}
\left\{-\frac{1}{2}\nabla^2 +v_\mathrm{s}(\mathbf{r})\right\}\phi_i(\mathbf{r}) = \epsilon_i^\tau\phi_i(\mathbf{r}) ~, \notag \\
n(\mathbf{r}) = \sum^N_{i=1} f_\mathrm{i} |\phi_i(\mathbf{r})|^2~, 
\end{gather} 

where $f_i$ is a temperature-dependent Fermi weighting. 
Solving the above equations generates the FT temperature density used to calculate $A^\tau[n]$, the total free energy of the system \cite{M65},

\begin{equation}
\label{eq:freeEnergy}
A^\tau[n] = T_\mathrm{s}[n] -\tau S_\mathrm{s}[n] + U_\mathrm{H}[n] +V_\mathrm{ext}[n] +A_\mathrm{xc}^\tau[n].
\end{equation}

\noindent Here, $\tau$ is temperature, $S_\mathrm{s}$ is the non-interacting entropy, and $A_\mathrm{xc}^\tau$ is the exchange-correlation free energy. 
The exchange-correlation free energy is given by

\begin{equation}
\label{eq:axc}
A_\mathrm{xc}^\tau[n] = T_\mathrm{xc}^\tau[n] - \tau S_\mathrm{xc}^\tau[n] + U_\mathrm{xc}^\tau[n]= K_\mathrm{xc}^\tau+U_\mathrm{xc}^\tau[n]~.
\end{equation} 

\noindent $U_\mathrm{xc}^\tau$ is the potential XC and $K_\mathrm{xc}^\tau$ is the kentropy, which is made up of the XC kinetic energy, $T_\mathrm{xc}^\tau$, and $\tau$ multiplied by the XC entropy, $S_\mathrm{xc}^\tau$. In practical use, $A_\mathrm{xc}^\tau$ must also be approximated, and a thermal ACF can help build these approximations. 
Some of the authors' previous work \cite{HMP22,HMXP24} has analyzed the temperature dependence of $A_\mathrm{xc}^\tau$ through the FTAC and GTAC for the uniform electron gas (UEG) model system. The FTAC extends the ground-state ACF to finite temperatures, and previous results show its successful application for the UEG at WDM conditions using an \textit{ab initio} parametrization for the exchange-correlation free energy from Groth and co-workers (GDSM) \cite{GDSM17}. 
This framework was later extended to the GTAC to allow for variations in both interaction strength and a fictitious temperature parameter, $\tau'$,

\begin{equation}
\label{eq:gtac_axc}
A_\mathrm{{xc}}^\tau[n] = E_\mathrm{{xc}}[n] + \int_{0}^{1} d\lambda \int_{0}^{\tau} d\tau' \frac{\partial}{\partial \tau'}\frac{U_\mathrm{{xc}}^{\tau',\lambda}[n]}{\lambda},
\end{equation}

\noindent where $E_\mathrm{{xc}}$ is evaluated on the finite-temperature density. It was shown that the exact exchange-correlation entropy,$S_\mathrm{xc}$, can be extracted\cite{HMXP24} from a Maxwell-style relationship \cite{BSGP15}:

\begin{equation}
\label{eq:maxwell}
\left(\frac{\partial U^{\tau,\lambda}_\mathrm{{xc}}\left[n\right]}{\partial \tau}\right)_\lambda = -\lambda\left( \frac{\partial S^{\tau,\lambda}_\mathrm{{xc}}\left[n\right]}{\partial \lambda}\right)_\tau  .
\end{equation}

\begin{figure}[ht]
\centering
\includegraphics[width=0.85\columnwidth]{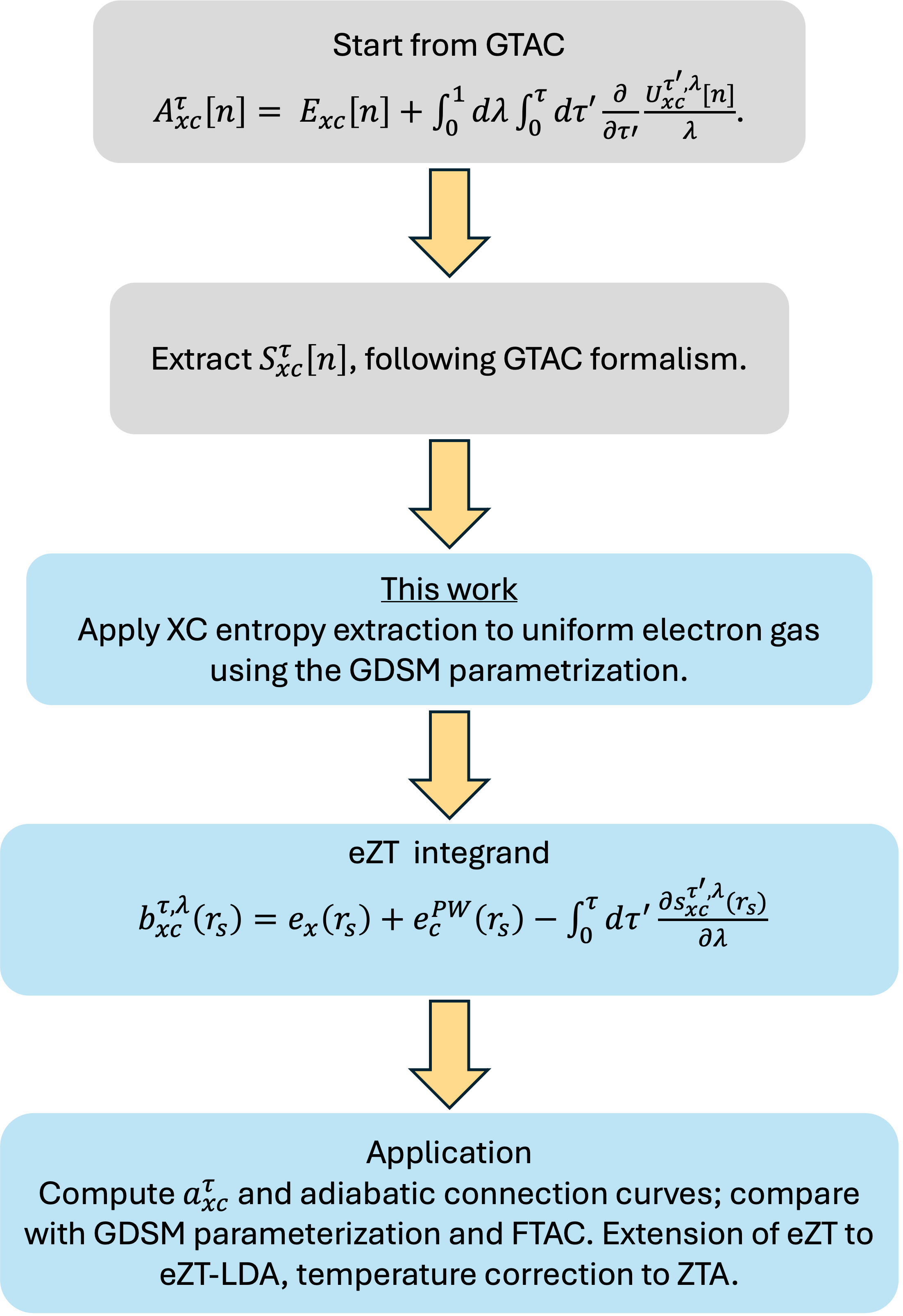}
\caption{\label{fig:workflow} Workflow for obtaining eZT adiabatic connection formula, starting from GTAC.}
\end{figure}

In this work, this relationship is applied to the UEG in building adiabatic connection curves, and extracting the XC free energy. The following derivation presents a thermal correction to the zero-temperature approximation for the UEG called the eZT approach, built by extracting the entropy following the method in Reference \cite{HMXP24}. The method explicitly uses the XC entropy to obtain the potential XC, in order to interpret the resulting adiabatic connection curves and the XC free energy. The workflow is depicted in Figure \ref{fig:workflow}.

\subsection{Derivation for eZT approach}
In this work, we formulate the eZT approach specifically for the UEG. As such, all energetic quantities are written in lowercase to denote energy per particle. While the methodology is general and can be extended to non-uniform systems, the UEG serves as a natural starting point, owing to its well-understood behavior and the availability of accurate parametrizations \cite{GDSM17,CA80}. This allows us to construct the approximation for the case in which it should agree with the exact case and allow for validation of the method. 

Starting with $E_\mathrm{xc}[n]$ in Equation (\ref{eq:gtac_axc}), the user may implement their favorite ground-state DFA, evaluated on a Fermi-weighted density. In this work, the exchange energy per particle is known exactly for the UEG in terms of the Wigner-Seitz radius, $r_\mathrm{s}$:

\begin{equation}
\label{eq:ex_ueg}
e_\mathrm{x}(r_\mathrm{s}) = \frac{-3}{4\pi}\left(\frac{9\pi}{4r_\mathrm{s}^3}\right)^{1/3}~.
\end{equation}

\noindent Alternatively, the correlation energy per particle for the UEG does not have an exact analytical form. In the work presented, we implement the Perdew-Wang (PW) parametrization for the ground-state correaltion energy per particle, $e_\mathrm{c}^\mathrm{PW}$ as it has been shown to perform well in the desired density regime \cite{PW92}. By choosing to use the PW ZT correlation, we've purposefully built a correction to the ZTA. This complements the framework of GTAC, where it has been shown that the XC entropy can be extracted from the GDSM parametrization./cite{HMXP24}    

For the second, $\tau$-dependent term in Equation (\ref{eq:gtac_axc}), we start from the GDSM parameterization of the XC free energy, $a_\mathrm{xc}^{\tau}$. We apply simulated scaling\cite{HMP22} for the interaction strength, $\lambda$, 

\begin{equation}
\label{lambda_axc}
    a_\mathrm{xc}^{\tau,\lambda}(n)=\lambda^2a_\mathrm{xc}^{\tau/\lambda^2}(\lambda r_\mathrm{s})~.
\end{equation}
Following the procedure in \cite{HMXP24}, we use a well-known thermodynamic relationship to extract the XC entropy from the $\lambda$-scaled, XC free-energy parameterization, 

\begin{equation}
\label{thermo_relationship}
    s_\mathrm{xc}^{\tau,\lambda}(r_\mathrm{s}) = -\frac{\partial a_\mathrm{xc}^{\tau,\lambda,\mathrm{GDSM}}(r_\mathrm{s})}{\partial \tau}~.
\end{equation}

\noindent Re-writing Equation (\ref{eq:maxwell}) and taking into account the simulated scaling in Equation (\ref{lambda_axc}),  we can obtain the adiabatic connection integrand from the XC entropy using, 

\begin{align}
\label{eq:uxclam_sxc}
    -\int_0^{\tau}d\tau'\frac{\partial}{\partial \lambda} \left( \lambda^2 s_\mathrm{xc}^{\tau'/\lambda^2}(\lambda r_\mathrm{s}) \right) & = \frac{\lambda^2u_\mathrm{xc}^{\tau'/\lambda^2}(\lambda r_\mathrm{s})}{\lambda}\bigg|^\tau_0 \notag \\
    & = \lambda u_\mathrm{xc}^{\tau'/\lambda^2}(\lambda r_\mathrm{s})\bigg|^\tau_0 \notag \\
    & = \lambda \left(u_\mathrm{xc}^{\tau/\lambda^2}(\lambda r_\mathrm{s}) - u_\mathrm{xc}^{0}(\lambda r_\mathrm{s}) \right) ~.
\end{align}

\noindent Combining all components, we construct the full adiabatic connection integrand for the UEG by explicitly going through the XC entropy. We denote this eZT integrand as $b_{xc}^{\tau,\lambda}$: 

\begin{equation}
\label{eq:b_xc}
b_\mathrm{xc}^{\tau,\lambda}(r_\mathrm{s}) = e_\mathrm{x}(r_\mathrm{s}) + e_\mathrm{c}^\mathrm{PW}(r_\mathrm{s})  -\int_0^{\tau}d\tau'\frac{\partial}{\partial \lambda} \left( \lambda^2 s_\mathrm{xc}^{\tau'/\lambda^2}(\lambda r_\mathrm{s}) \right)~.
\end{equation}

\noindent Integrating $b_\mathrm{xc}^{\tau,\lambda}$ over $0\leq \lambda \leq 1$ yields the exchange-correlation free energy for the uniform electron gas:

\begin{equation}
\label{eq:b_xc_int}
a_\mathrm{xc}^{\tau}(r_\mathrm{s}) = \int_0^1d\lambda ~ b_\mathrm{xc}^{\tau,\lambda}(r_\mathrm{s})~. 
\end{equation}

In the following section, this formulation is applied to the UEG, across a range of WDM conditions, to evaluate $a_\mathrm{xc}^\tau$ and compare the results with the GDSM parametrization. The corresponding adiabatic connection curves are also examined in relation to the FTAC formulation.

\section{\label{results} Results \& Discussion}

The eZT approach was applied across varying Wigner-Seitz radii, $r_\mathrm{s}$, and temperatures, $\tau$, that would yield an electron degeneracy within $0.5 \leq \Theta \leq 8$ and $0.1 \leq r_\mathrm{s} \leq 20$, to be consistent with the fitting range of the GDSM parametrization \cite{GDSM17}. Note that $\Theta$ is related to the Wigner-Seitz radius via $\Theta = \left(\frac{4\sqrt{8}}{9\pi}\right)^{2/3}\tau r_\mathrm{s}^2$. Calculations were completed in Mathematica version 14.0 \cite{Mathematica}. The geometric interpretation of the adiabatic connection curves from the eZT approach, their comparison to FTAC, and performance are described in the following sections.

\subsection{Geometric interpretation}

\begin{figure}[ht]
\centering
\includegraphics[width= 1\columnwidth]{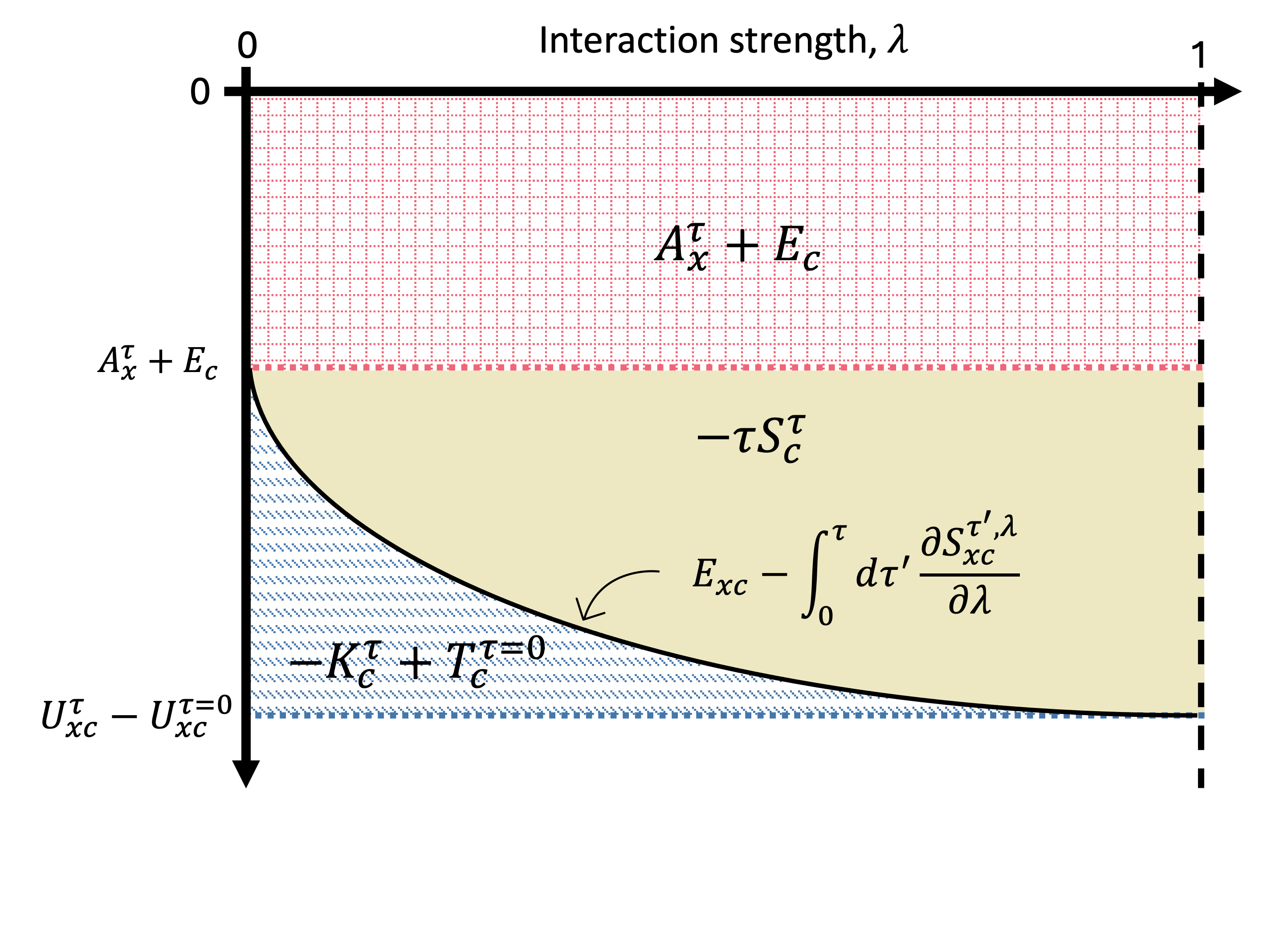}
\caption{\label{fig:geom_interp}Geometric interpretation of eZT adiabatic connection, general case.}
\end{figure}

The geometric interpretation of the adiabatic connection is illustrated in Figure \ref{fig:geom_interp}. The depiction presented here is general and serves to illustrate the relationship between $b_\mathrm{xc}^{\tau,\lambda}$ and $a_\mathrm{xc}^\tau$ before applying the framework to the uniform electron gas. The following derivation shows how the XC free energy is recovered from the eZT adiabatic connection integrand, by integrating over $\lambda$:

\begin{align}
\label{eq:recover_axc_sxc}
A_\mathrm{xc}^{\tau}[n] & =\int_0^1 d\lambda \left(E_\mathrm{xc}[n] - \int_0^\tau d\tau ' \frac{\partial S_\mathrm{xc}^{\tau ',\lambda}[n]}{\partial \lambda}\right) \notag \\
&= E_\mathrm{xc}[n]  - \int_0^\tau d\tau ' S_\mathrm{xc}^{\tau '}[n] \notag \\
&= E_\mathrm{xc}[n] - \left(\tau S_\mathrm{xc}^{\tau ' = \tau}[n] - \tau S_\mathrm{xc}^{\tau ' = 0}[n]\right) \notag \\
&= E_\mathrm{xc}[n]- \tau S_\mathrm{xc}^\tau[n] .
\end{align}

This leads to the geometric interpretation of the eZT shown in Figure \ref{fig:geom_interp}. The starting point of the curve corresponds to the sum of the exchange free energy and the correlation energy. As such, the area above the curve to the starting point corresponds to the temperature multiplied by the correlation entropy. From this, we recover the complete exchange-correlation free energy as expected. In effect, the eZT adiabatic connection curve itself (shown in black in Figure \ref{fig:geom_interp}) captures the $\lambda$ dependence of the components of the XC free energy that depend on {\em both} temperature $\tau$ and interaction strength $\lambda$.

When the entropy term of the integrand is left in terms of the XC potential, the equivalent of integrating equation (\ref{eq:gtac_axc}) over $\lambda$, exposes the relationship between FTAC and eZT: 

\begin{align}
\label{eq:recover_axc_uxc}
A_\mathrm{xc}^{\tau} & =\int_0^1 d\lambda \left(E_\mathrm{xc} + \int_0^\tau d\tau ' \frac{\partial}{\partial \tau'} \frac{U_\mathrm{xc}^{\tau ',\lambda}}{\lambda}\right) \notag \\
&= E_\mathrm{xc} + \int^0_1 d\lambda \left(\frac{U_\mathrm{xc}^{\tau,\lambda}}{\lambda}-\frac{U_\mathrm{xc}^{\tau=0,\lambda}}{\lambda}\right) \notag \\
&= E_\mathrm{xc}+ A_\mathrm{xc}^{\tau}-E_\mathrm{xc}  .
\end{align}

Essentially, the FTAC is recovered when evaluating the GTAC at finite temperature, as expected. Moreover, when viewed geometrically, the eZT can be understood as the difference between the FTAC and the zero-temperature ACF. This relationship is reflected in the total areas highlighted in Figure \ref{fig:geom_interp}. The area under the eZT curve corresponds to the negative of the temperature-dependent correlation kentropy, $K_\mathrm{c}^\tau=T_\mathrm{c}^\tau-\tau S_\mathrm{c}^\tau$, added to the zero-temperature kinetic correlation. In other words, the eZT geometric interpretation effectively represents the subtraction of the finite- and zero-temperature ACF areas, consistent with the FTAC geometric interpretation shown in Figure \ref{fig:ftac_updated}. In this way, the eZT approach naturally lends itself to an LDA-like temperature correction to the zero-temperature XC energy, since it explicitly removes the zero-temperature contribution from the total XC free energy. 

\subsection{Adiabatic connection curves}

\begin{figure}[ht]
\centering
\includegraphics[width= 1\columnwidth]{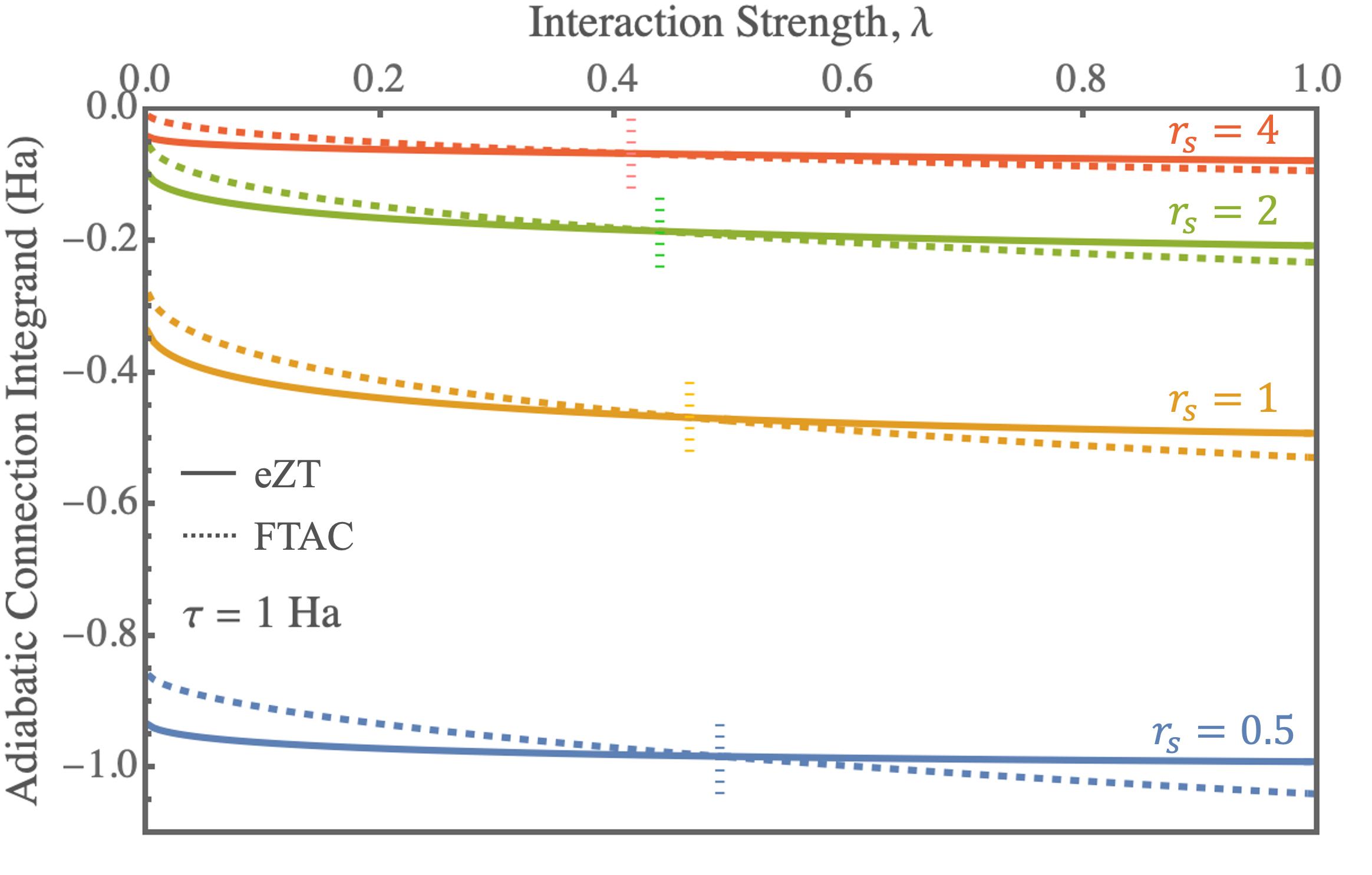}
\caption{\label{fig:intersection_rs} AC curves for FTAC (dashed) and eZT(solid) at $\tau = 1$ Ha for various $r_\mathrm{s}$ (Bohr) values, note intersection point denoted by vertical lines at intersection points.}
\end{figure}

\begin{figure}[ht]
\centering
\includegraphics[width= 1\columnwidth]{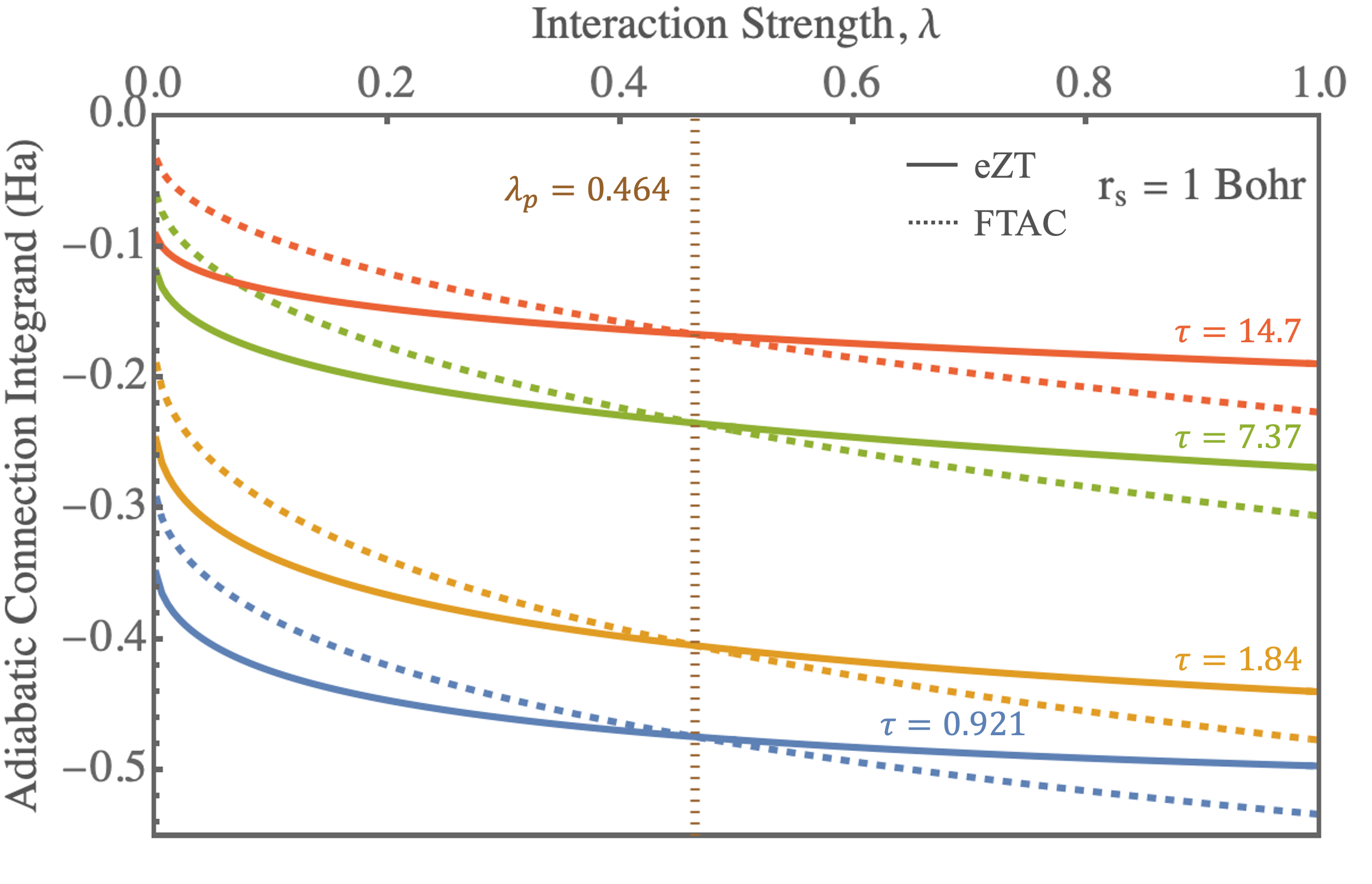}
\caption{\label{fig:intersection_theta} AC curves for FTAC (dashed) and eZT(solid) at $r_\mathrm{s} = 1$ for various $\Theta$, intersection point at $\lambda_\mathrm{p}$ = 0.464 for all eZT and FTAC curves.}
\end{figure}

 A comparison of the eZT and FTAC adiabatic connection curves at varying $r_\mathrm{s}$ and $\Theta$ are shown in Figures \ref{fig:intersection_rs} and \ref{fig:intersection_theta}. The plots for the eZT adiabatic connection differ in both the starting point and curvature from the FTAC. The downward shift of the eZT curves at $\lambda = 0$ is expected, seen in that the zero-temperature correlation energy is lambda-independent in the described formalism and therefore constant across the range $\lambda=[0,1]$. As a result, the difference in the FTAC and eZT starting point is equivalent to $E_\mathrm{c}^\mathrm{DFA}$, which in this work is $e_\mathrm{c}^\mathrm{PW}$. This is a manifestation of the different starting point for the eZT, made up of $a_\mathrm{x}^\tau$ and $e_\mathrm{c}^\mathrm{DFA}$, as opposed to just the exchange free energy in the FTAC. 

\begin{figure}[ht]
\centering
\includegraphics[width= 1\columnwidth]{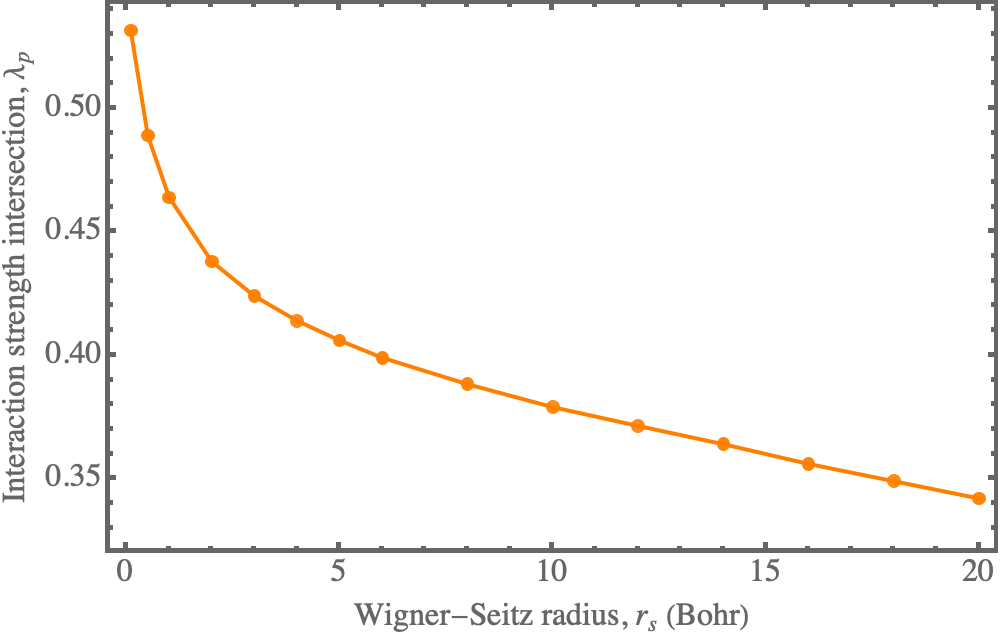}
\caption{\label{fig:intersection} Interaction strength, $\lambda_p$, intersection point of the FTAC and eZT curves for each sampled $r_\mathrm{s}$ value. }
\end{figure}

We also note a point of intersection between the entropy-corrected and FTAC schemes, denoted in Figures \ref{fig:intersection_rs} and \ref{fig:intersection_theta} as vertical lines. The latter plot shows that the intersection point is temperature-independent, while the former shows the intersection point changes with $r_\mathrm{s}$. Figure \ref{fig:intersection} shows the intersection point, $\lambda=\lambda_p$, decreases with increasing $r_\mathrm{s}$. 

In order to further investigate this intersection point, we first separate the exchange and correlation free energies by incorporating simulated scaling of the coordinates, $\gamma$, and scaling to the high-density limit, 

\begin{equation}
\label{eq:lim_axc}
a_\mathrm{x}^{\tau,\mathrm{GDSM}}(r_\mathrm{s}) = \lim_{\gamma \rightarrow \infty} \frac{a_\mathrm{xc}^{\gamma^2 \tau, \mathrm{GDSM}} \left(\frac{r_\mathrm{s}}{\gamma}\right)}{\gamma}~.
\end{equation}

The correlation free-energy is obtained by taking the difference of the total exchange-correlation free energy and the exchange free energy:

\begin{equation}
\label{eq:ac}
a_\mathrm{c}^{\tau, \mathrm{GDSM}} (r_\mathrm{s}) =   a_\mathrm{xc}^{\tau,\mathrm{GDSM}}(r_\mathrm{s}) - a_\mathrm{x}^{\tau, \mathrm{GDSM}}(r_\mathrm{s})~. 
\end{equation}

The corresponding entropy components are obtained following the eZT derivation above, leaving exchange and correlation separated. At this intersection point, $\lambda_\mathrm{p}$, the adiabatic connection integrand for eZT (shown below with XC separated,left) is equal to that of FTAC (right),

\begin{align}
\label{eq:intersection_equality}
e_\mathrm{x}(r_\mathrm{s}) - \tau s^\tau_\mathrm{x}(r_\mathrm{s})+ 
& e_\mathrm{c}^{PW}(r_\mathrm{s})-\int_0^{\tau}d\tau'\frac{\partial}{\partial \lambda}\left. \left( \lambda^2 s_\mathrm{c}^{\tau'/\lambda^2}(\lambda r_\mathrm{s}) \right)\right\vert_{\lambda=\lambda_\mathrm{p}} \notag \\
&= a_\mathrm{x}^\tau (r_\mathrm{s}) + \lambda_\mathrm{p} u_\mathrm{c}^{(\tau/\lambda_\mathrm{p}^2)}(\lambda_\mathrm{p} r_\mathrm{s})~.
\end{align}

Given that $a_\mathrm{x}^\tau(r_\mathrm{s}) = e_\mathrm{x}(r_\mathrm{s}) - \tau s^\tau_\mathrm{x}(r_\mathrm{s})$, the exchange free energy cancels and we're left with the correlation components. Re-writing the correlation entropy to be in terms of the correlation potential, using Equation (\ref{eq:uxclam_sxc}), we obtain:

\begin{equation}
\label{eq:intersection_correlation}
e_\mathrm{c}^{PW}(r_\mathrm{s}) + \lambda_\mathrm{p} \left(u_\mathrm{xc}^{\tau/\lambda_\mathrm{p}^2}(\lambda_\mathrm{p} r_\mathrm{s}) - u_\mathrm{xc}^{0}(\lambda_\mathrm{p} r_\mathrm{s}) \right)  = \lambda_\mathrm{p} u_\mathrm{c}^{(\tau/\lambda_\mathrm{p}^2)}(\lambda_\mathrm{p} r_\mathrm{s})~.
\end{equation}

\noindent Simplifying the equation to combine like terms and solving for the intersection, $\lambda_\mathrm{p}$ yields:

\begin{equation}
\label{eq:intersection_solved}
\lambda_\mathrm{p} u_\mathrm{c}^{0}\left( \lambda_\mathrm{p} r_\mathrm{s}\right) = e_\mathrm{c}^{PW}\left(  r_\mathrm{s}\right) ~.
\end{equation}

The eZT-FTAC intersection point, $\lambda_\mathrm{p}$, is temperature-independent and relies only on the ground-state correlation potential and correlation energy. At $\lambda_\mathrm{p}$, the ZT limit of the correlation-only FTAC integrand equals the average value of the ZT integrand, which in turn corresponds to the ground state correlation energy. As we expect, the entropic correlation components of the eZT are completely offset by the temperature-dependent components of the correlation potential in the FTAC integrand.


\begin{center}
\begin{table*}[t]%
\caption{Table of values for GDSM XC free energies (Ha) and eZT- XC free energies (Ha) within $0.5 \leq \Theta \leq 8$.\label{tab:axc_theta}}
\begin{tabular*}{\textwidth}{@{\extracolsep\fill}lllllllll@{}}
&\multicolumn{2}{c}{$\boldsymbol{\Theta = 0.5}$} & \multicolumn{2}{c}{$\boldsymbol{\Theta = 1}$}& \multicolumn{2}{c}{$\boldsymbol{\Theta = 4}$}& \multicolumn{2}{c}{$\boldsymbol{\Theta = 8}$} \\\cmidrule{2-3}\cmidrule{4-5}\cmidrule{6-7}\cmidrule{8-9}
\textbf{$\mathbf{r_s}$ (Bohr)} 
& {GDSM $a_\mathrm{xc}$}  
& {eZT $a_\mathrm{xc}$}    
& {GDSM $a_\mathrm{xc}$}  
& {eZT $a_\mathrm{xc}$}
& {GDSM $a_\mathrm{xc}$}  
& {eZT $a_\mathrm{xc}$}  
& {GDSM $a_\mathrm{xc}$}  
& {eZT $a_\mathrm{xc}$}  
 \\
\midrule
0.1 & -3.51317 & -3.52256 & -2.58651 & -2.59600 & -1.11772 & -1.12734  & -0.70805  & -0.71769  \\
1   &  -0.46598 & -0.46682  &  -0.39506 & -0.39591 &  -0.22846 & -0.22933 & -0.16271 & -0.16358  \\
4   & -0.14247 & -0.14222 & -0.13063 & -0.13039 & -0.09039 & -0.09016 & -0.06958 & -0.06935  \\
20  &  -0.03451 & -0.03441 & -0.03355 & -0.03346 & -0.02792 & -0.02782 & 
-0.02370 & -0.02361   \\
\bottomrule
\end{tabular*}
\end{table*}
\end{center}


\begin{table*}[t]
\centering
\caption{Relative error (in Ha) for XC free energies within $0.5 \leq \Theta \leq 8$.}
\label{tab:axc_error}
\begin{tabular*}{\textwidth}{@{\extracolsep\fill}lcccc@{}}
\toprule
\textbf{$\mathbf{r_s}$ (Bohr)} & $\boldsymbol{\Theta = 0.5}$ & $\boldsymbol{\Theta = 1}$ & $\boldsymbol{\Theta = 4}$ & $\boldsymbol{\Theta = 8}$ \\
\midrule
0.1 & -0.0027 & -0.0037 & -0.0086 & -0.0136 \\
1   & -0.0018 & -0.0021 & -0.0038 & -0.0053 \\
4   & -0.0017 & -0.0018 & -0.0026 & -0.0034 \\
20  & -0.0028 & -0.0028 & -0.0034 & -0.0040 \\
\bottomrule
\end{tabular*}
\end{table*}

We can also investigate how the curves depend on temperature and verify that they follow known conditions. Previous work \cite{HMP22} found that while $A_\mathrm{c}$ changes non-monotonically with temperature, the magnitude of $A_\mathrm{x}$ decreases with increasing temperature. Our results for the eZT align with these prior findings. Figure \ref{fig:intersection_theta} depicts AC curves at $r_\mathrm{s} = 1$ and shows that the area above the curve decreases as $\tau$ increases. This area corresponds to $a_\mathrm{x}^\tau + e_\mathrm{c} $. From this, we know that the decrease in magnitude comes from $a_\mathrm{x}^\tau$ as this is the only temperature-dependent piece in this area. 

\subsection{Performance of eZT}

\begin{figure}[ht]
\centering
\includegraphics[width=1\columnwidth]{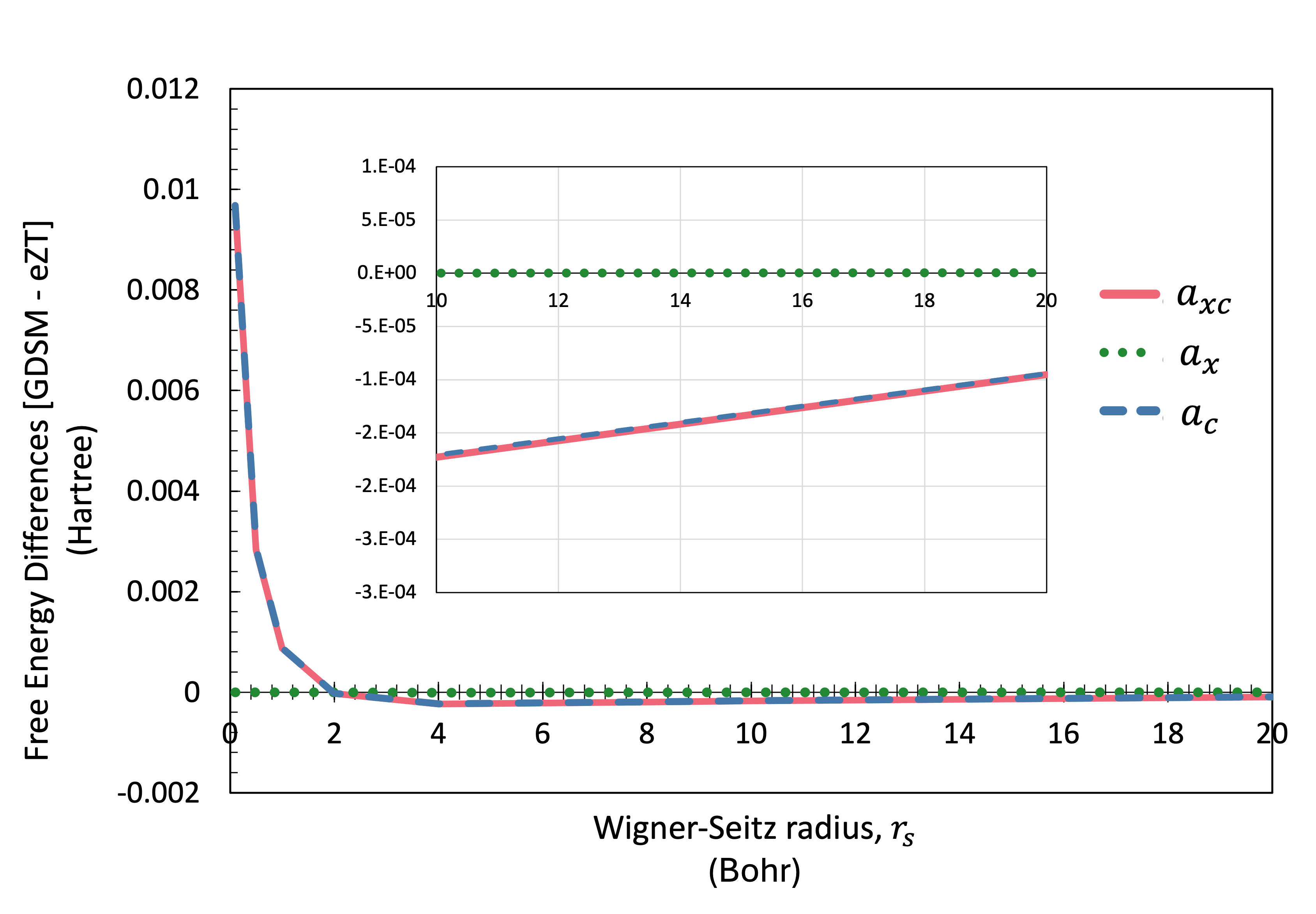}
\caption{\label{fig:FreeEDiff}Free energy difference between GDSM parametrization and eZT for exchange-correlation (solid,pink), exchange(dotted, green), and correlation (dashed,blue) free energies at $\Theta = 8.$}
\end{figure}

To analyze the performance of the eZT approach, the exchange and correlation contributions are separated using the scaling relationship shown in Equation (\ref{eq:lim_axc}). 
We observe in Figure \ref{fig:FreeEDiff}, the exchange free energy obtained from eZT is consistent with that of the GDSM parametrization, across all Wigner-Seitz radii considered. In the eZT approach, the total exchange contribution is evaluated by integrating with respect to $\tau'$ to obtain $-\tau s_\mathrm{x}(r_\mathrm{s})$.  

As shown in Figure \ref{fig:FreeEDiff}, all observed differences are contained in the correlation component. This behavior is attributed to the use of the PW parametrization for the ground state correlation energy in the eZT approach. Differences between PW and the GDSM correlation, evaluated at near zero temperature ($\tau = 0.0001$ Ha), match the difference observed in the eZT approach versus GDSM. The PW parametrization was constructed for $0.5 \leq r_\mathrm{s} \leq 100$, therefore it follows that the greatest difference, 0.01 Ha, is observed at $r_\mathrm{s} = 0.1$ Bohr. 

Tables \ref{tab:axc_theta} and \ref{tab:axc_error} compare the XC free energies obtained from GDSM and eZT methods at various electron degeneracies. The largest relative error in $a_\mathrm{xc}^\tau$ occurs at $r_\mathrm{s}=0.1$ and $
\Theta=8$, the edge of the GDSM parametrization. For a given Wigner-Seitz radius, the largest error also occurs at $\Theta=8$, the edge of the GDSM parametrization. Despite the larger deviations observed at $\Theta = 8$, the overall mean absolute error for $a_\mathrm{xc}$, considering all sampled $r_s$ and $\Theta$ values, is  0.0027 Ha ($\sim$1.69 kcal/mol), demonstrating that the eZT method reproduces GDSM $a_\mathrm{xc}$ values quite accurately in most cases within the parametrization range of applicability. 

\section{\label{conclusion}Conclusion}

The eZT approach to the generalized thermal adiabatic connection, based on the exchange-correlation entropy, explicitly accounts for entropic temperature dependence. Our results for the UEG show good agreement with the Groth parametrization obtained from quantum Monte Carlo data in the WDM regime. Thus far, the framework has been applied specifically to the UEG, noting its foundational role in XC functional approximations and usefulness as a well-understood electronic system. In this model system, the eZT approach accurately reproduces the XC free energy, with the highest accuracy occurring as $r_s$ grows. This may provide a powerful benefit when applied under further approximations below, where eZT is combined with standard ZT DFAs that are typically more reliable at higher densities. Furthermore, these low-$r_s$ errors are shown to be based on the differences between the PW parametrization for ground-state correlation and the GDSM values, inviting speculation that improved parameterizations may provide even higher accuracy in these high-density regions. 

Additionally, the eZT and FTAC adiabatic connection integrands intersect at a well-defined interaction strength, $\lambda_{p}$, that is independent of temperature but varies with density. By separating exchange and correlation contributions, we show that the exchange terms cancel at the intersection, leaving a dependence on the correlation components, specifically the potential and energy per particle. This result shows how the zero-temperature limit of the $\lambda$-scaled correlation potential is equivalent to the average value of the ZT integrand. 

Extension of the eZT approach beyond model systems is feasible by formulating an approximate form of the eZT. In particular, rewriting the eZT in terms of the density, $n$, rather than the Wigner-Seitz radius, enables us to construct a density functional approximation. Work is currently underway to implement the eZT in the form of a local density approximation--like correction to the ground-state exchange-correlation energy evaluated on the Fermi-weighted density (sometimes called the ``zero-temperature approximation"). This scheme captures the XC entropy at an LDA level, while maintaining the advantages of higher-level DFAs for the terms exhibiting implicit temperature dependence in simulations at non-zero temperatures. The general form of the eZT-LDA density functional takes in the Fermi-weighted density,  

\begin{equation}\label{eq:gen_LDA}
A^{\tau,{\rm eZT-LDA}}_\mathrm{xc}[n^\tau] = E^{\rm DFA}_\mathrm{xc}[n^\tau] - \tau S^{\tau,{\rm eZT-LDA}}_\mathrm{xc}[n^\tau],
\end{equation}
where
\begin{equation}
   S^{\tau,{\rm eZT-LDA}}_\mathrm{xc}[n^\tau]=\int d^3r ~ n({\bf r})s_\mathrm{xc}^\mathrm{unif}\left(n({\bf r})\right) 
\end{equation}
\noindent and $s_\mathrm{xc}^\mathrm{unif}$ is the expression for the UEG XC entropy per particle extracted in this work.  In this way, the eZT approach can be leveraged, albeit imperfectly, in systems for which we do not have the XC entropy at hand. Once implemented, this approximation will be evaluated for different DFAs for the zero-temperature approximation, to evaluate which ZT approximations work best in combination with the eZT LDA-like correction. 

Future investigations will explore the performance of the eZT-LDA correction against existing thermal XC functionals, such as corrKSDT \cite{KSDT14,KVDT18}, to assess its performance across different density and temperature regimes. Additionally, comparisons with experimental observables, like x-ray Thomson scattering measurements \cite{wunsch_ion_2009,schorner_x-ray_2023,bohme_correlation_2026}, could help identify where entropy effects become most critical, which can in turn guide the refinement of thermalized DFAs. 

Overall, the eZT outlined here provides a pathway to remedy the missing temperature dependence in zero-temperature approximations. By explicitly including these effects, we anticipate more accurate temperature-dependent simulations, with applications ranging from astrophysics to inertial confinement fusion experiments. 

\section{Acknowledgments}
We thank Professor Hrant P. Hratchian and Dr. Vincent Martinetto for useful discussions. This work was supported by the National Science Foundation Graduate Research Fellowship Program under Grant No. 2139297 and the Department of Energy under Award No. DE-SC0024476. Calculations were conducted using Pinnacles (NSF MRI, No. 2019144) at the Cyberinfrastructure and Research Technologies (CIRT) at University of California, Merced.

\section{Conflict of interest}
The authors declare no potential conflict of interests.

\section{Supporting information}

The data that supports the results are available from the corresponding author upon request. A Mathematica notebook featuring the derivations and calculations in this paper is available at \url{https://www.hypugaea.com/publications}.

\bibliography{references}

\end{document}